\documentclass[11pt,a4paper]{article}
\usepackage[margin=2.5cm]{geometry}
\usepackage[utf8]{inputenc}
\usepackage{authblk}
\usepackage{graphicx}
\usepackage{subcaption}
\usepackage{hyperref}
\usepackage{amsmath}
\usepackage{amssymb}
\usepackage{slashed}
\usepackage{color}
\usepackage{booktabs}
\usepackage[shortlabels]{enumitem}
\usepackage[nameinlink,capitalize]{cleveref}
\usepackage{cite}
\usepackage{bbold}

\newcommand{\uone}{U(1)_\mathrm{FN}}
\newcommand{\Lfn}{\Lambda_\mathrm{FN}}
\newcommand{\Lew}{\Lambda_\mathrm{EW}}

\title{\textbf{Reviving keV sterile Neutrino Dark Matter}}
\author{Carlos Jaramillo\footnote{email: carlos.jaramillo@mpi-hd.mpg.de}}
\affil{\textit{Max-Planck-Institut f\"ur Kernphysik, Saupfercheckweg 1, 69117 Heidelberg, Germany}}
\date{}

\begin{document}

\maketitle

{\small {\flushleft \url{https://doi.org/10.1088/1475-7516/2022/10/093} \hfill  Carlos Jaramillo JCAP10(2022)093}}

\begin{abstract}\noindent
We propose a new production mechanism for keV sterile neutrino dark matter which relies neither on the oscillations between sterile and active neutrinos nor on the decay of additional heavier particles. The dark matter neutrinos are instead produced by thermal freeze-out, much like a typical WIMP. The challenge consists in balancing a large Yukawa coupling so that the sterile neutrinos thermalize in the early universe on the one hand, and a small enough Yukawa coupling such that they are stable on cosmological scales on the other. We solve this problem by implementing varying Yukawa couplings. We achieve this by using a three-sterile neutrino seesaw extension to the SM and embedding it in a Froggatt-Nielsen model with a single flavon. Because the vev of the flavon changes during the electroweak phase transition, the effective Yukawa couplings of the fermions have different values before and after the phase transition, thus allowing for successful dark matter genesis. Additionally, the hierarchy in the flavour structure is alleviated and the origin of the light neutrino masses is explained by the interplay of the seesaw and Froggatt-Nielsen mechanisms.
\end{abstract}

\section{Introduction}
The phenomenon of neutrino flavour oscillations, which has been observed \cite{Fukuda_1998, SNO:2001kpb,SNO-2002} and confirmed by many experiments \cite{Bellini:2013wra,Wang:2015rma} and for which the 2015 Nobel Prize was awarded \cite{RevModPhys.88.030501}, implies definitively that active neutrinos have tiny non-vanishing masses.
Also, there is an overwhelming amount of evidence for the existence of a non-baryonic substance, which makes up almost one quarter of the energy density of the Universe today and is generally referred to as Dark Matter (DM) \cite{Kolb:1990vq,Dodelson2003,Profumo-book-2015,bertone-book,Planck2018}.
It is widely accepted that the origin of the light neutrino masses and the fundamental nature of the Dark Matter of the Universe are evidence of the incompleteness of the Standard Model (SM) of particle physics as a theory of nature.
Perhaps the simplest and most elegant solution to the problem of neutrino masses is delivered by the well known seesaw mechanism of type I \cite{Minkowski:1977sc,Yanagida:1979as,GellMann:1980vs,Mohapatra:1980yp,Schechter:1980gr}.
Therein, at least two right-handed (RH) SM singlet fermions are introduced, which have a Majorana mass term and couple to the SM neutrinos through a Yukawa coupling, thus repairing the left-right asymmetry in the matter content of the theory; these are the right-handed (also called sterile) neutrinos. 
Just like in the case of all other fermions, the neutrinos get a Dirac mass term through the Higgs mechanism. 
The full neutrino-mass-matrix contains the Majorana mass matrix and the Dirac mass matrix.
After diagonalization, and assuming that the elements of the Majorana matrix are much larger than those of the Dirac matrix, the active-neutrino mass matrix arises and is suppressed by the large entries in the Majorana matrix, thus elegantly explaining the smallness of the masses of the active neutrinos.
This minimal extension of the SM is also highly appealing because one (or many) of the sterile neutrinos could play the role of the DM of the Universe (see for instance $\nu$MSM \cite{Shaposhnikov2005}).
There are two basic requirements for this to be the case: firstly, there has to be a viable mechanism to produce the right density of sterile neutrinos in the early universe, i.e.\,the observed relic abundance.
Secondly, the sterile neutrinos must be stable on cosmological timescales, i.e.\,they must be compatible with the indirect (non-) detection observations.
This last issue is non-trivial because the sterile neutrinos, being massive and coupled to the active neutrinos by the Yukawa term, may decay into SM particles.
Given the results of the indirect detection searches, it has turned notoriously difficult to reconcile these two requirements with each other: the most straightforward way to produce sterile neutrino DM in the early Universe is through the oscillations with active neutrinos, the so-called "Dodelson-Widrow mechanism" \cite{Dodelson1994}.
However, this simplest scenario has already been ruled out by indirect search observations \cite{Perez:2016tcq,Baur:2017stq} 
because the Dodelson-Widrow mechanism requires a sizeable mixing between active and sterile neutrinos; however, if one allows for large enough mixing angles, the decay rate of the DM neutrinos increases, thus conflicting with indirect detection results.
This constrains can be somewhat alleviated if the oscillations that produce the DM occur resonantly (the so-called Shy-Fuller mechanism \cite{Shi_Fuller1998}) but this requires the presence of a lepton asymmetry at temperatures well below the Electroweak (EW) scale, which is hard to motivate. 
Another possibility is that sterile neutrinos are produced by the decay of additional dark-sector heavy particles, see for example \cite{Kusenko:2006rh,Petraki2008,Kusenko:2010ik,Kusenko2009,Alanne2018,Bezrukov:2017ike,Datta:2021elq}, or that the production is aided by the presence of non-standard neutrino interactions \cite{Hansen:2017rxr, DeGouvea:2019wpf, Kelly:2020pcy, Sen:2021mxl, Benso:2021hhh, Bringmann:2022aim}.
However, such models depart from the minimal seesaw extension to the SM and in them the active-sterile coupling plays little to no role at all, thus rendering indirect searches less relevant.

Therefore, an alternative mechanism for sterile neutrino DM genesis on the basis of the Yukawa coupling is desirable.
We obtained our inspiration from another very popular candidate as a DM particle, a generic Weakly Interacting Massive Particle (WIMP).
WIMPs are coupled to the SM by some portal with an interaction strength similar or smaller than the weak interactions of the SM.
Thanks to such interactions WIMPs are generally thought to be in thermal equilibrium with the primordial plasma until their interaction rate falls below the expansion rate of the Universe as it expands and cools, at which point the comoving density of the WIMPs gets frozen-out and remains constant thereafter.
This production mechanism is usually referred to as ``freeze-out", and until now it has not been considered for sterile neutrinos because the Yukawa coupling necessary to thermalize them would also make them highly unstable - they would not survive long.

Here we propose a scenario where this problem is solved by implementing varying Yukawa couplings.
The main idea is simple: if the neutrino Yukawa coupling is allowed to be large enough in the early universe for the sterile neutrinos to thermalize, but later drops to much smaller values such that their interaction rate turns smaller that the expansion rate of the Universe, then the relic abundance of sterile neutrinos will be fixed by a process similar to the WIMP freeze-out.
Varying Yukawa couplings have been studied in the past, particularly in the context of EW Baryogenesis \cite{Servant:2018xcs,vonHarling:2016vhf,Baldes:2016gaf}.
There are many ways to implement varying Yukawa couplings and, in principle, any of them could work to thermalize sterile neutrinos in the early Universe and then induce their freeze-out to generate the observed DM relic abundance.
For concreteness, however, we will examine the implementation of varying Yukawa couplings involving the Froggatt-Nielsen  mechanism.
The Froggatt-Nielsen (FN) mechanism is a popular way to explain the flavour structure in the SM (i.e.\,the mass hierarchy in the fermion sector).
It employs a new scalar field, called the flavon, and an additional $U(1)$ flavour symmetry under which the flavon and the fermions are charged.
The bare Yukawa couplings, which within this framework were all initially of order 1, become effectively suppressed by the vev of the flavon in a hierarchical pattern determined by the different flavour charges of the fermions.
In this work we consider the possibility that the flavon vev changes during the Electroweak Phase Transition (EWPT), varying from a value similar to the flavour scale before the EWPT to a somewhat smaller value afterwards.
This is a reasonable consideration because the scalar potential responsible for the EWPT will contain both the Higgs and the flavon fields and therefore it will find its minimum in the Higgs-flavon field space.
Thus, during the EWPT, as the potential relaxes to its true minimum, one should not expect the coordinate of the new minimum along the flavon axis to remain constant.
On the contrary, it is sensible to expect that the flavon-coordinate of the minimum will also change as the Higgs-coordinate changes.
As a consequence, the Yukawa couplings will have different effective values before and after the phase transition.
As we will see, something similar to this will also happen to the Majorana masses of the sterile neutrinos.

Here we will formulate a concrete low energy FN model which explains neutrino masses and oscillations via the seesaw mechanism as well as the flavour hierarchy in the lepton sector via the FN mechanism.
Moreover, we will show that the variation in the flavon vev leads to effectively varying Yukawa couplings that allow for the thermalization of the sterile neutrinos before the EWPT and their freeze-out afterwards, such that the lightest sterile neutrino with keV mass plays the role of the DM and its relic abundance is set by thermal freeze-out.

The outline of this paper is as follows.
We begin in \cref{sec:seesaw} by reviewing the seesaw mechanism and briefly discussing the keV neutrino as a DM candidate.
Then, in \cref{sec:dm-genesis} we introduce the DM production mechanism and compute the DM relic abundance.
In \cref{sec:fn-model} we formulate a low energy FN model to implement the varying Yukawa couplings for DM production, the light neutrino masses and the lepton flavour hierarchy, before stating some concluding remarks in \cref{sec:conclusio}.

\section{The keV sterile neutrino as the Dark Matter particle}
\label{sec:seesaw} 

We start by reviewing the general Type I seesaw mechanism.
The SM is extended by three\footnote{The low energy neutrino data can be explained within the type I seesaw framework with two neutrinos or more. However, anticipating \cref{sec:fn-model} we discuss the case of three sterile neutrinos, which is also the case in the well known $\nu$MSM.} neutral Majorana fermions that are also singlets with respect to the SM symmetry, $(\nu_R)_i$. 
With the presence of these sterile neutrinos the Lagrangian gets the following additional terms:
\begin{align}\label{eq:L-neutrino}
-{\cal L} \supset i \bar{\nu}_R\slashed{\partial}\nu_R + \bar{L}\, Y_\nu \, \Tilde{\phi}\nu_R + \frac{1}{2} \overline{\nu^c_R} \, M_R \, \nu_R  + \mathrm{h.c.},
\end{align}
where $\tilde \phi$ is the dual Higgs field, $L$ stands for the $SU(2)$ lepton doublets, $Y_\nu$ is the neutrino Yukawa matrix and $M_R$ is the Majorana mass matrix.
After the spontaneous breaking of the EW symmetry by the vev of the Higgs $v$, a Dirac mass matrix for the neutrinos is generated with $m_D = Y_\nu\,v/\sqrt{2}$.
The mass term for all six neutrinos can then be formulated as
\begin{align}
    - {\cal L}_\mathrm{\nu, mass} = \frac{1}{2}\,\overline{\nu^c_{\cal M}}\, {\cal M}_\nu \, \nu_{\cal M} + \mathrm{h.c.} =  \frac{1}{2}\,\overline{\nu^c_{\cal M}}
    \begin{pmatrix}
    0 & m_D \\
    m_D^T & M_R
  \end{pmatrix} \nu_{\cal M} + \mathrm{h.c.},
\end{align}
where the basis is defined by $\nu_{\cal M} = (\nu_{L,e},\nu_{L,\mu},\nu_{L,\tau}, \nu_{R,1}^c, \nu_{R,2}^c, \nu_{R,3}^c)^T$ and ${\cal M}_\nu$ is a $6\times6$ matrix.
The full neutrino mass matrix can first be block diagonalized to
\begin{align}
{\cal M}_\nu' = \mathrm{diag}(m_\nu, \,M_N)
\end{align}
and, assuming that the seesaw condition holds, namely that $m_D\ll M$, we obtain
\begin{align}\label{eq:seesaw-non-d}
	m_\nu \approx m_D\,M^{-1}\,m_D^T, \qquad M_N \approx M_R ,
\end{align}
whereby these matrices can be further diagonalized by the appropriate matrices $V$ and $U$
\begin{align}
    m_\nu^d = V^T \, m_\nu \, V = \mathrm{diag}(m_1,m_2,m_3), \qquad M_N^d = U^T\, M_N\, U = \mathrm{diag}(M_1,M_2,M_3),
\end{align}
resulting in three very light neutrino mass eigenstates $\nu_i$, mostly composed of the active flavours, and three heavy states $N_i$, mostly composed of the sterile neutrinos.
The active-sterile mixing is then of the order $\theta\sim m_D\,M_R^{-1}$.

It is the lightest sterile neutrino $N_1$ which we want to consider as the DM particle.
The so-called \textit{Tremaine-Gunn Bound} \cite{tremaine1979dynamical} states that a fermionic DM particle candidate must be heavier than $\sim 1\,\mathrm{keV}$. This model independent constrain is based on the fact that the densest phase space distributions for fermionic DM particles may not be larger than that allowed by the Pauli exclusion principle.
Thus, sterile neutrinos with masses in the keV scale are the lightest viable fermionic DM candidates.
They are also appealing because they would be warm dark matter (WDM), meaning that they tend to suppress the power of structure formation at small scales, offering a possible solution to the satellite problem and the core-cusp problem \cite{Core-CuspProblem,missing-satellites}.

Regardless of how they are produced, sterile neutrinos with a Yukawa coupling to the SM as in \cref{eq:L-neutrino} are never completely stable.
The interaction strength/probability of $N_1$ with the SM will be characterised by the sum of the square of its mixing angle with all active neutrinos \cite{Boyarsky:2009ix}, i.e.
\begin{align}\label{eq:theta12}
 \theta_1^2 = \sum_{\alpha = e,\mu,\tau} \theta_{\alpha, 1}^2= \sum_{\alpha = e,\mu,\tau} \frac{v^2 \, |(Y_\nu)_{\alpha 1}|^2}{M_1^2} = \sum_{\alpha = e,\mu,\tau} \frac{|(m_D)_{\alpha 1}|^2}{M_1^2}.
\end{align}
Through this mixing, sterile neutrinos with keV masses can decay at tree level to three active neutrinos, $N_1\rightarrow \nu_i\,\bar \nu_j\,\nu_j$.
The fact that DM must still exist today means that the lifetime of unstable DM should be comparable to the age of the universe, delivering a constraining relationship between $M_1$ and $\theta_1$.
However, an even more constraining relation is obtained by the sub-dominant one-loop decay $N_1 \rightarrow \nu_i\,\gamma$, where the produced photon is almost monochromatic with energy $E_\gamma = M_1/2$.
These photons, which for $M_1 \sim \mathrm{keV}$ will lie in the $X$-Ray part of the spectrum, can be searched for in space with $X$-Ray telescopes \cite{Dolgov:2000ew,Abazajian:2001vt,denHerder:2009sxr}.
The non-observation of this smoking-gun signal has placed a limit on the mixing angle of sterile neutrino DM \cite{Drewes2017},
\begin{align}\label{eq:xraybound}
\sin^2(2\theta_1) \lesssim 3 \times 10^{-5} \left(\frac{\mathrm{keV}}{M_1}\right)^5,
\end{align} 
assuming that $N_1$ makes up 100\% of the DM abundance.

Since the year 2014 there has been a lot of debate concerning a suspicious $X$-Ray line with an energy of $E_\gamma = 3.55\,\mathrm{keV}$. The signal has been detected from a multitude of sources, such as galaxy clusters, the Andromeda galaxy and our own galactic center, and a multitude of instruments e.g.\,\textit{XMM-Newton, Chandra, EPIC, Suzaku, ACIS} and \textit{Fermi}, just to name a few.
The reader may find a review in \cite{Iakubovskyi:2015wma}.
This signal can be interpreted as the smoking-gun photon from sterile neutrino DM decay, implying a DM mass of $M_1=7.1\,\mathrm{keV}$.
While this is a very attractive hypothesis and there have been many arguments made in its favour \cite{Drewes2017}, it remains controversial.
The line could also originate from astrophysical processes \cite{Jeltema:2014qfa} or appear as a consequence of unaccounted systematics \cite{Boyarsky:2014jta}.
Particularly, the line has not been observed from dwarf spheroidal galaxies, which were expected to provide the cleanest DM decay signal because of their high mass-to-light ratio and correspondingly low $X$-Ray background \cite{Malyshev:2014xqa}.
The debate for the correct interpretation of this signal is still ongoing and one can only hope that more data and further analysis will bring clarity.

In any case, if we insist on the DM interpretation of the signal, an alternative to the Dodelson-Widrow mechanism for DM production is needed, because the mixing angle necessary to explain the relic abundance by the Dodelson-Widrow mechanism is incompatible with the value that would correspond to the 3.55\,keV line.
Although resonant production is still allowed, it would require an additional mechanism to generate a large lepton asymmetry well after the EWPT, which is not well motivated.
To address this situation, we propose in the following section an alternative production mechanism that provides an explanation to the 3.55 keV signal that is compatible with the DM hypothesis.

\section{Sterile neutrino Dark Matter genesis by freeze-out}
\label{sec:dm-genesis}

To simultaneously explain the phenomena of active neutrino oscillations and to offer a viable DM candidate, three sterile neutrinos are needed: one to play the role of the DM and the other two of them to generate the masses of the light neutrinos.
Thus, we consider an extension to the SM by three sterile neutrinos.
In this work we will insist that the relic abundance of the lightest sterile neutrino $N_1$, which shall play the role of the DM particle, is produced in the early Universe by thermal freeze-out, similar to a typical WIMP, and not by oscillations from active neutrinos or by the decay of additional heavier particles.
For this to work, $N_1$ must have a large enough Yukawa coupling so that it comes into thermal equilibrium with the SM bath.
On the other hand, a large Yukawa coupling translates into a large active-sterile mixing angle $\theta_1$, which would lead to a rapid decay of the DM particles and contribute a sharp line at $E=M_1/2$ to the $X$-Ray background and from sources with a large DM density or mass to light ratio. 
These problems are solved if, by some mechanism, the Yukawa coupling varies during the early universe, going from large values at early times, to smaller values at later times.
Even though there are many ways to implement varying Yukawa couplings \cite{Servant:2018xcs}, we here chose to do so by embedding the seesaw extension in a Froggatt-Nielsen model.
This has the added benefit that it also helps to explains the flavour hierarchy.

\begin{table}[tbp]
\begin{center}
\begin{tabular}{lccccc}
\toprule
    Field    & $L_{i}$  & $E_{R_j}$ & $N_k$ & $\Theta$ & $\phi$\\
\midrule  
    $U(1)_\mathrm{FN}$ Charge &  $q_{L_i}$   &  $q_{R_j}$   &  $q_{N_k}$ & $-1$ & 0  \\
\bottomrule
\end{tabular}
\caption{\label{tab:FNcharges} FN charges for the lepton doublets $L_i$, charged singlets $E_{R_j}$, sterile neutrinos $N_k$, the flavon $\Theta$ and the Higgs $\phi$, respectively.}
\end{center}
\end{table} 
The Froggatt-Nielsen (FN) mechanism \cite{Froggatt:1978nt} attempts to resolve the flavour hierarchy puzzle by introducing a scalar field, called the flavon\footnote{The flavon encodes the Froggatt-Nielsen messengers, which are heavier particles that are integrated out below the flavour scale $\Lfn$.} $\Theta$, and a global flavour symmetry $\uone$, under which the flavon and all fermions are charged.
The flavon is set to have a $\uone$ charge of $-1$, while the fermions $f_i$ may have any integer as their charge $q_{f_i}$ and the Higgs field $\phi$ is not charged under the flavour symmetry, see \cref{tab:FNcharges}. 
Consequently, the Yukawa terms in the Lagrangian arise from non-renormalizable operators that contain powers of the flavon such that the terms are invariant under the $\uone$ symmetry. 
Thus, below the flavour scale $\Lfn$ the usual Yukawa terms in the SM Lagrangian must be replaced by
\begin{align}\label{eq:FN-yukawa-term}
    Y_{ij}\, \overline{f_L}_i\,\phi \, f_{R\,j} \longrightarrow  Y_{ij}\,\overline{f_L}_i\,\phi \,f_{R\,j}\, \left(\frac{\langle \Theta \rangle}{\Lfn}\right)^{q_{L,i}\, + \, q_{R,j}}.
\end{align}
The key idea is that below the flavour scale $\Lfn$ the flavon has a non-vanishing vev, and therefore each entry in the Yukawa and Majorana matrices gets rescaled by powers of the FN factor, defined as $\lambda := \langle \Theta \rangle/\Lfn$.
This resolves the flavour hierarchy puzzle because, if $\lambda <1$, then each entry is effectively suppressed by powers of $\lambda$ according to the sum of FN charges of the fermions involved.
Thus, the observed flavour hierarchy is the result of the different FN charges of the fermions and not due to a mysterious hierarchy in the Yukawa couplings, which in this framework are all allowed to be of order unity.
This mechanism is usually employed in the quark sector, where the FN suppression factor is set to $\lambda \approx 0.22$, in resemblance of the Cabibbo angle.
However, in this work we will apply these tools to the lepton sector to study the posibility of sterile neutrino DM.
For the lepton sector, the relevant terms in the effective Lagrangian are
\begin{align}\label{eq:seesawL}
-{\cal L_\mathrm{eff}} &\supset 
	\bar{L_i}\,Y^E_{ij}\,{\phi}\,E_{R\,j} \,\left( \frac{\langle\Theta\rangle}{\Lfn} \right) ^{q_{\bar L i}+q_{R j}}
	 + \, \bar{L_i}\,Y^\nu_{ij}\,\Tilde{\phi}\,\nu_{R j}\,\left( \frac{\langle\Theta\rangle}{\Lfn} \right) ^{q_{\bar L i}+q_{N j}} \\
	&+ \, \frac{1}{2} \overline{\nu_{R i}^c} \, (M_R)_{ij} \,\nu_{R j}\,\left( \frac{\langle\Theta\rangle}{\Lfn} \right)^{q_{N i}+q_{N j}} \,+\, \mathrm{h.c.}, \nonumber
\end{align}
where $q_{L,R,N}$ are the $\uone$ charges of the $SU(2)$ charged doublets, singlets and sterile neutrinos respectively, as given in \cref{tab:FNcharges}, and all fermions should be understood as containing all three generations.

\begin{figure}[bt]
\begin{center}
	\includegraphics[width=0.9\textwidth]{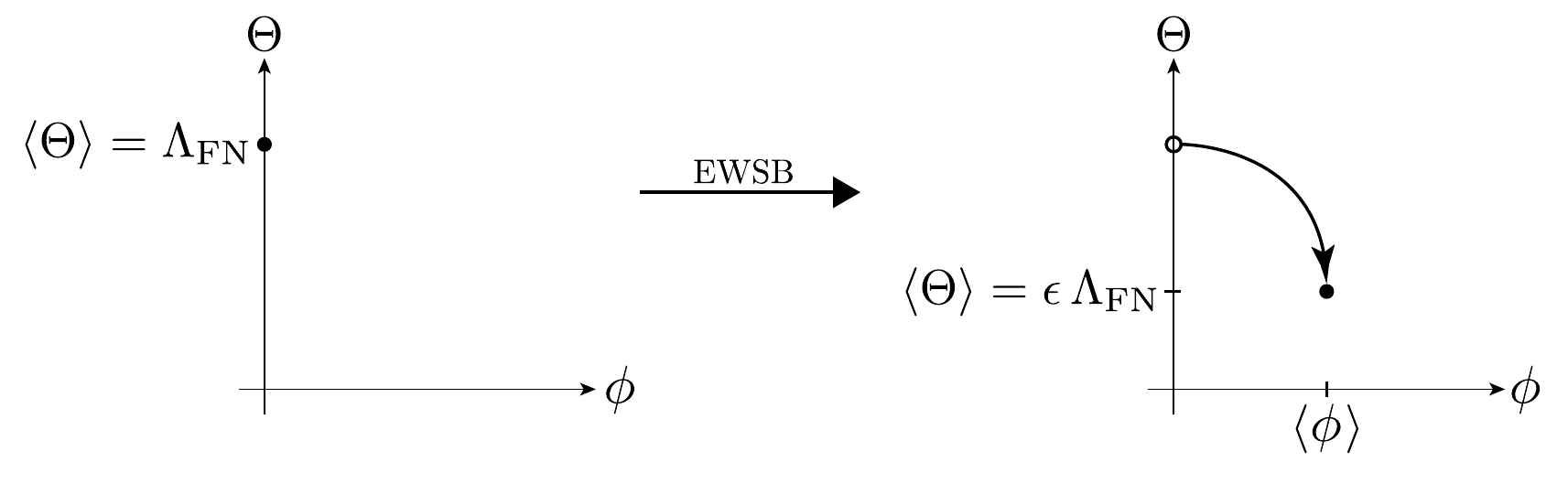}
	\caption{Location of the minimum of the scalar potential in field space and shift in the value of the flavon vev caused by the electroweak symmetry breaking (EWSB).}
	\label{fig:phi-pt}
\end{center}
\end{figure}
To achieve the thermal freeze-out of sterile neutrinos, we propose to implement varying Yukawa couplings in the following way, resembling \cite{Baldes:2016gaf,Jaramillo_2021}.
Since the scalar potential of the theory will include both the Higgs $\phi$ and the flavon $\Theta$, the true minimum of the scalar potential will be located at a point in the $(\phi,\Theta)$-field space.
Below the flavour scale but prior to the EWPT (i.e.\,at a temperature $T$ with $\Lew<T<\Lfn$), the vev of the Higgs is $\langle\phi\rangle=v=0$ while the vev of the flavon is $\langle\Theta\rangle = w_e \neq 0$ and has already broken the flavour symmetry (the subscript $e$ stands for \textit{early}, as in $w_e$ is the vev of the flavon at times earlier than the EWPT).
If we set $ w_e = \Lfn$, then the FN factor would be $\lambda = 1$ at early times and the Yukawa couplings of order unity would not be suppressed.
Crucially, this means that the sterile neutrinos would be in thermal equilibrium because their Yukawa couplings would be as strong as that of the top quark.
Then, as the Universe cools down, the EWPT kicks in and plays a key role: as the Higgs gets its vev and the scalar potential relaxes to its true minimum, it is reasonable to expect that this minimum lies at a completely new position in the $(\phi,\Theta)$-field space, or in other words, during the EWPT the vev of the flavon gets dragged along by the vev of the Higgs.
This is illustrated in \cref{fig:phi-pt}.
If after the EWPT the vev of the flavon ends up at a value $\langle\Theta\rangle = w_l = \epsilon \, \Lfn$ with $\epsilon<1$, then the Yukawa couplings get suppressed by powers of $\epsilon$ (here, the subscript $l$ stands for \textit{late}, as in $w_l$ is the vev of the flavon at times later than the EWPT).
This suppression could be drastic enough to force the sterile neutrinos to decouple from thermal equilibrium and could even make the lightest sterile neutrino stable on cosmological time scales and thus evade the $X$-Ray bounds.
In this context, the FN suppression factor $\lambda$ can be understood as describing a path in field space with boundaries given by
\begin{align}\label{eq:FN-factor}
	\lambda(\langle \phi \rangle) =
	\begin{cases}
	1,\quad & \mathrm{for} \quad \langle \phi \rangle = 0 \\
	\epsilon,\quad & \mathrm{for} \quad \langle \phi \rangle = v.
	\end{cases}
\end{align}
The precise trajectory in field space during the EWPT is not relevant.
What matters is that the flavon vev has different values in the different phases, as sketched in \cref{fig:phi-pt}.
We will comment on the scalar potential necessary to achieve this behaviour at the end of this section.

Note from \cref{eq:seesawL} that this mechanism not only suppresses the Yukawa matrices but also the Majorana mass matrix is affected.
Just like the Yukawa matrices, the Majorana mass matrix can be understood as effectively having different values before and after the EWPT.
As we will discuss below, this issue is highly significant for DM genesis.
In terms of the $\lambda$ parameter, and recalling \cref{eq:seesaw-non-d}, we can write the $(j,k)$ elements of the effective sterile neutrino mass matrix $M_\mathrm{eff}$ as
\begin{align}\label{eq:initial_final_M}
	(M_\mathrm{eff})_{j k} = (M_N)_{j k} \, [\lambda(\langle \phi \rangle)]^{q_{N j} \,+\, q_{N k}} =
		\begin{cases}
			(\tilde M)_{j k} = (M_N)_{j k}, &\mathrm{for}\quad  T>T_{EW} \\
			(M)_{j k} = (M_N)_{j k} \,\epsilon^{q_{N j} \,+\, q_{N k}}, &\mathrm{for}\quad T<T_{EW} 
		\end{cases},
\end{align}
where $\tilde M$ stands for the \textit{early} Majorana mass matrix, i.e.\,before it is suppressed by the shift in the flavon vev at the EWPT and $M$ is the \textit{late} Majorana mass matrix, i.e.\,the suppressed Majorana mass matrix after the EWPT.
Diagonalising the $\tilde M$ and $M$ gives us the masses of the sterile neutrinos before and after the EWPT respectively. 
We will refer to the early (i.e.\,unsuppressed) and late (i.e.\,suppressed) mass of the lightest sterile neutrino, i.e.\,the DM neutrino, as $\tilde M_{1}$ and $M_1$ respectively.

We can convince ourselves that the sterile neutrinos were indeed in thermal equilibrium before being forced to decouple by the EWPT by computing the ratio of the sterile neutrino interaction rate $ \Gamma_\nu $ to the expansion rate of the Universe $ H(T) $.
Prior the EWPT, the main interactions keeping the sterile neutrinos in thermal equilibrium are the decays and inverse decays governed by the Yukawa couplings, which are all $ \mathcal{O}(1) $ since they have not been suppressed yet.
The corresponding interaction rate and the Hubble rate during radiation domination are given by
\begin{align}
  \Gamma_\nu = \frac{(Y^\nu)^2 \tilde M_1}{16\,\pi}, \qquad H(T) \approx \frac{\sqrt{g_\rho(T)}}{M_\mathrm{P}}\,T^2,
\end{align}
where $ g_\rho(T) $ stands for the effective number of energetic d.o.f. in the plasma at temperature $ T $, $ M_\mathrm{P} $ is the reduced Planck mass and we have ignored the fact that the Yukawa coupling is actually a matrix, just for simplicity.
Inserting the values for $T =\Lambda_\mathrm{EW}$, $ \tilde M_1 = 10^3\,$GeV and taking $ Y^\nu\sim \mathcal{O}(1) $ we arrive at
\begin{align}
	\left. \frac{\Gamma_\nu}{H} \right|_{T\gtrsim T_\mathrm{EW}} \sim 10^{14} \gg 1,
\end{align}
thus confirming that prior to the suppression induced by the EWPT, the sterile neutrinos are very much in thermal equilibrium.
However, after the EWPT the Yukawa couplings will be suppressed by a factor of $ \epsilon^{q_{\bar L}+q_{N}} $.
Assuming that $\epsilon = 0.1$, as we will do throughout this paper, this implies that the sum of the FN charges of the $ SU(2) $ leptons and sterile neutrinos should be at least 7 in order for the suppression to force the sterile neutrinos to decouple, i.e. $ q_{\bar L}+q_{N} > 7 $.
In fact, the concrete example which will be studied in the next section, will be such that $ q_{\bar L}+q_{N} = 11 $, leading to 
\begin{align}
	\left. \frac{\Gamma_\nu}{H} \right|_{T\lesssim T_\mathrm{EW}} \sim 10^{-8} \ll 1,
\end{align}
 which is more than enough of a suppression to decouple the sterile neutrinos (the choice of $ q_{\bar L}+q_{N} = 11 $ is made in order to also avoid the stringent $X$-ray bounds, as will be discussed in the next section).
 Furthermore, the fact that the EWPT also causes a suppression of the Majorana mass matrix will lead to a further reduction of the interaction rate.
Particularly, if the suppressed mass is lighter than the Higgs mass, i.e.\,$ M_1<m_\phi $, as would be the case for $ M_1\sim\mathcal{O(\mathrm{keV})} $, then the direct decay of the sterile neutrinos becomes kinematically forbidden.
Then, the only interactions the sterile neutrinos can have with the SM are through active-sterile oscillations, which will be negligible because of the  tiny mixing angle, which is proportional to the Yukawa coupling, $ \theta_{\alpha i} = Y^\nu_{\alpha i}\,v/M_i $.
Indeed, although further production (or depletion) of sterile neutrinos through the Dodelson-Widrow mechanism (active-sterile oscillations) is in principle still possible, the drastically suppressed mixing angle renders this effect  negligible \cite{Merle2016}. 
 
The knowledge that the suppression of the Yukawa couplings is enough to force the sterile neutrinos to drop out of thermal equilibrium (i.e.\,to freeze-out) is now enough to compute the relic abundance of the lightest sterile neutrino as described in the following.
The contribution to the total energy density of the Universe from the frozen-out DM neutrinos, which are non-relativistic today, is given by \cite{Kolb1990}
\begin{align}\label{eq:relic-abundance}
	\Omega_\mathrm{DM} \, h^2 = \frac{s_0\,y_\mathrm{fo}\,M_1 }{\rho_\mathrm{crit}/h^2},
\end{align}
where $s_0$ is the entropy density today, $y_\mathrm{fo}$ is the frozen-out comoving number density of the sterile neutrinos, $\rho_\mathrm{crit}$ is the critical density of the Universe and the Hubble parameter today is given by $H_0 = 100\,h \; \mathrm{km\,s}^{-1} \mathrm{Mpc}^{-1}$.
The early masses of the sterile neutrinos will be of the same order as the Majorana scale, which is expected to be higher than $\Lew$.
Thus, shortly before the EWPT the sterile neutrinos with early mass $\tilde M_1$ will be non-relativistic and their comoving density in equilibrium will be given by
\begin{align}\label{eq:y_eq}
y_\mathrm{eq}(T,\tilde M_1)=\frac{45}{(2\pi^5)^{3/2}g_s}\left( \frac{\tilde M_1}{T} \right)^{3/2}\,e^{-\tilde M_1/T},
\end{align}
where $g_s$ is the effective entropic number of degrees of freedom in the plasma.
Then, when the EWPT occurs and shifts the vev of the flavon, the drastically suppressed Yukawa couplings force the sterile neutrinos to drop out of thermal equilibrium and their comoving density is frozen at $y_\mathrm{fo} = y_\mathrm{eq}(T_{EW},\tilde M_1)$.
Notice that while it is the early mass $\tilde M_1$ that comes into $y_\mathrm{fo}$ in \cref{eq:y_eq}, the relic abundance today is proportional to the late mass $M_1$ in \cref{eq:relic-abundance}, because the energy density of non-relativistic DM today is given by number density of DM particles $n_\mathrm{DM}=s_0\cdot y_\mathrm{fo}$ times their mass today, i.e.\,the late mass $M_1$.
Thus, the relic abundance of our DM sterile neutrinos in this framework as given in \cref{eq:relic-abundance} can be understood as a function of three parameters, namely the freeze-out temperature, which here is equal to the temperature of the EWPT, $T_\mathrm{fo}=T_\mathrm{EW}$ and the early and late mass of the DM neutrinos $\tilde M_1$ and $M_1$.
\begin{figure}[tb]
\begin{center}
	\includegraphics[width=1\textwidth]{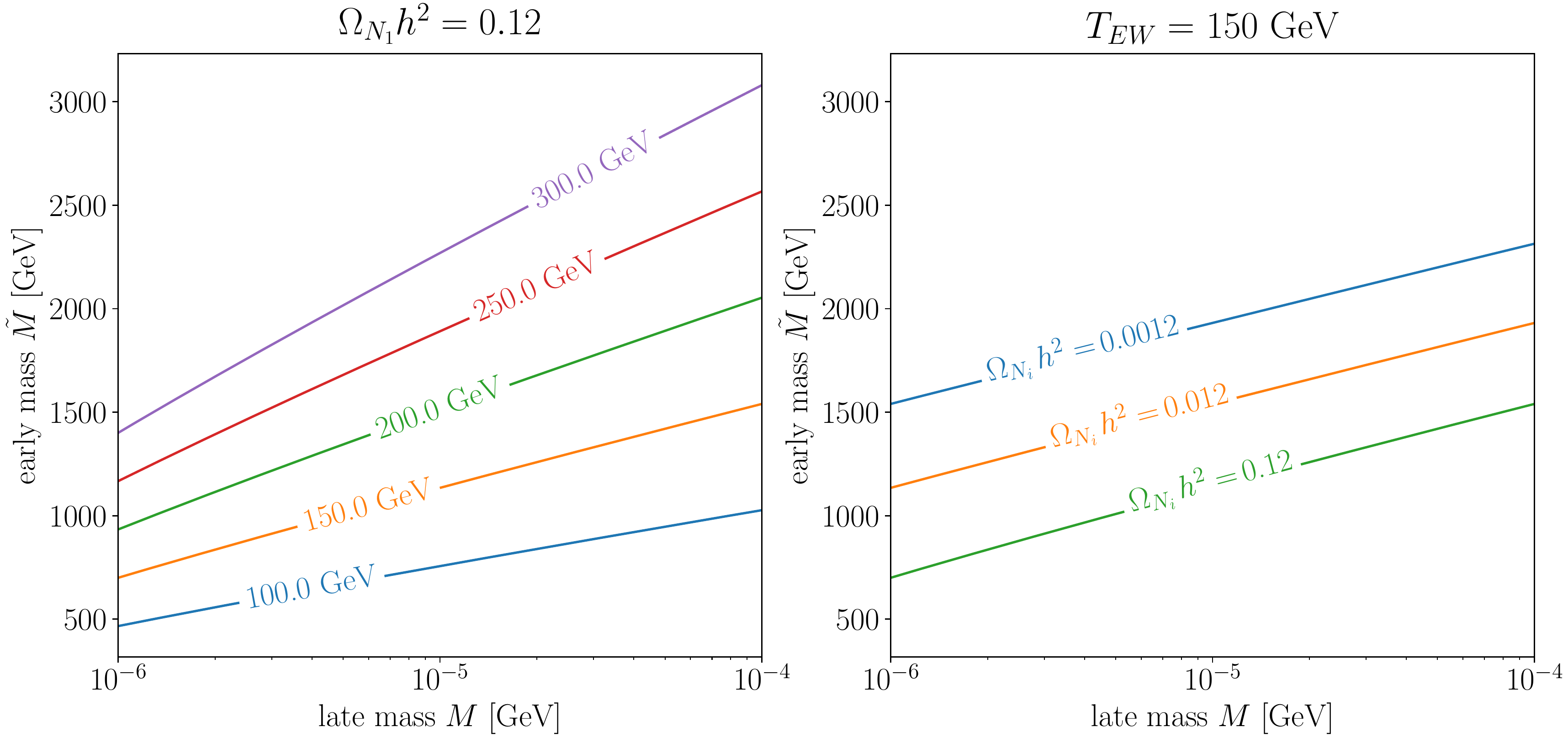}
	\caption{Left panel: lines for which the combinations of early and late masses result in a 100\% contribution from the sterile neutrinos to the DM abundance of the Universe for different freeze-out temperatures $T_\mathrm{fo}=T_{EW}$. For example, with a late mass of $M_1=7.1\,$keV and at a decoupling temperature of $T_\mathrm{fo}=150\,$GeV, one can see that if the early mass is $\tilde M_1\approx1\,$TeV, then the sterile neutrino can account for 100\% of the DM density. Right panel: At a fixed freeze-out temperature of $T_\mathrm{fo}=T_{EW}=150\,$GeV, the lines show the combinations of early and late masses for which the sterile neutrinos contribute 100\%, 10\% and 1\% to the DM abundance. Above the uppermost line, the contribution can be considered negligible.}
	\label{fig:full-abundance-lines}
\end{center}
\end{figure}

Using the values from reference \cite{Zyla:2020zbs} for $\Omega_\mathrm{DM} \, h^2$, $s_0$ and $\rho_\mathrm{crit}$, we can plot \cref{eq:relic-abundance} in the $(\tilde M_i, M_i)$ plane.
The results can be seen in the left panel of \cref{fig:full-abundance-lines}.
Each line represents a combination of suppressed and unsuppressed masses such that \cref{eq:relic-abundance} returns the full DM abundance with the freeze-out temperature $T_\mathrm{fo}=T_{EW}$ as an additional parameter.
For example, at $T_{EW}=150\,$GeV and $M_1=7.1\,$keV, the early DM mass should be $\tilde M_1\approx1\,$TeV for $N_1$ to account for all of the DM of the Universe.
Similarly, the right panel of \cref{fig:full-abundance-lines} depicts the relic abundance of sterile neutrinos for a fixed freeze-out temperature.
For instance, with $T_{EW}=150\,$GeV, if the two heavier sterile neutrinos have a mass of $M_2=10\,$keV and $M_3=50\,$keV after the EWPT and a mass of $\tilde M_2=2\,$TeV and $\tilde M_3=3\,$TeV before, then their contribution to the DM abundance will be negligible.
It is easy to understand why this is the case: when the induced freeze-out at $T_{EW}$ generates the DM abundance from the $N_1$ population, the population of $N_{2,3}$ is already depleted by the Boltzmann factor $\exp(-M_{2,3}/T_{EW})$ because of their larger masses $M_{2,3}>M_1$. 

It is thus clear that successful DM production by the mechanism proposed in this work depends only on three parameters: the temperature of the EWPT and the early and late masses of the sterile neutrinos.

As has been discussed in \cite{Jaramillo_2021,Baldes:2016gaf} \footnote{In \cite{Jaramillo_2021}, the PT during which the flavon vev changes is not necessarily the EWPT, but the PT of some generic scalar field $ \Sigma $ which undergoes SSB. To be applied in case in question here, one simply has to replace $ \Sigma $ by the Higgs field $ \phi $. The potential discussed in \cite{Jaramillo_2021} was taken and adapted from \cite{Baldes:2016gaf}.}, a tree-level, zero temperature scalar potential capable of exhibiting the desired behaviour, i.e.\,a simultaneous change in the vev's of both the Higgs and the flavon, should feature negligible Higgs-flavon mixing and could be of the form
\begin{align}
  V(\phi,\,\Theta) = \mu_\phi^2 \,\phi^\dagger\phi + \lambda_\phi \,(\phi^\dagger\phi)^2 + 
				  \mu_\Theta^2\,\Theta^\dagger \Theta + \lambda_\Theta \,(\Theta^\dagger\Theta)^2 + \lambda_{\phi\, \Theta}(\Theta^\dagger\Theta)(\phi^\dagger\phi),
\end{align}
where $ \lambda_\phi,\;\lambda_\Theta,\;\lambda_{\phi\,\Theta} $ are the self couplings of the Higgs, the flavon and the Higgs-flavon coupling, respectively. 
These couplings should not be confused with the FN suppression factor $ \lambda = \langle \Theta \rangle / \Lfn $ from \eqref{eq:FN-factor}.
The temperature and quantum correction to such a potential with the desired behaviour during the phase transition are discussed in \cite{Baldes:2016rqn}.
We point out that the simultaneous variation in the Higgs and flavon vev's is actually not really crucial for the DM production mechanism proposed here: the shift in the flavon vev could also occur before or after the EWPT; whenever it occurs, that will be the moment in which the sterile neutrinos are forced to decouple.
Such two-stepped phase transitions, in which the two scalar fields develop their vev's one after the other, have been considered e.g.\,in \cite{Blasi:2022woz}.  
The assumption that the shift in the flavon vev is realised during the EWPT simply allows us to fix the freeze-out temperature to the temperature of the EWPT, i.e.\,$ T_\mathrm{fo} = T_\mathrm{EW} $.
A more detailed study concerning the phenomenology of the scalar potential and the phase transition itself, however, lies beyond the scope of this work.

\section{Towards a model with a keV neutrino as Dark Matter, active neutrino masses and lepton flavour hierarchy}
\label{sec:fn-model}

Now we attempt to construct a low energy FN model that accomplishes the following goals:
\begin{enumerate}[label=(\roman*)]
	\item generate the light masses and mixing of the active neutrinos in accordance with experimental data through the type I seesaw mechanism.
	\item alleviate the flavour hierarchy in the lepton sector by exploiting the synergy of the seesaw and FN mechanisms.
	\item provide a DM candidate by implementing varying Yukawa couplings in the early universe to allow sterile neutrinos to be in thermal equilibrium and then freeze-out into long-lived stability, as described in the previous section.
\end{enumerate}

In other words, the task is to specify the FN charges of all leptons so that, with ${\cal O}(1)$ elements in all Yukawa matrices, the above conditions are satisfied.
The relevant part of the Lagrangian is given in \cref{eq:seesawL}.
We postulate that the FN suppression factor $\lambda$ has the following values before and after the EWPT:
\begin{align}
	\lambda = \frac{\langle \Theta \rangle}{\Lfn} =
	\begin{cases}
		1 &:\quad \text{before EWPT}\\
		\epsilon = 0.1 &:\quad \text{after EWPT}
	\end{cases}.
\end{align} 
Our starting point is the mass Lagrangian for the lepton sector in the broken EW phase:
\begin{align}\label{eq:L_m}
	-\mathcal{L}_m = \overline{E_{L \alpha}}\,(m_E)_{\alpha \beta}\,E_{R\,\beta} + \overline{ \nu_{L \alpha}} \, (m_D)_{\alpha i}\,\nu_{R i} + \frac{1}{2}\,\overline{\nu_{R i}^c}\, (M)_{i j}\,\nu_{R j}, 
\end{align} 
\begin{align}\label{eq:mass-matrices-in-fn}
	(m_E)_{\alpha \beta} = v \, \epsilon^{q_{\bar L_\alpha}}\,Y^E _{\alpha \beta}\,\epsilon^{q_{E_\beta}}, \quad (m_D)_{\alpha i} = v\,\epsilon^{q_{\bar L_\alpha}}\, Y^\nu_{\alpha i}\,\epsilon^{q_{N_i}}, \quad (M)_{i j} = \epsilon^{q_{N_i}}\,(M_N)_{i j}\,\epsilon^{q_{N_i}},
\end{align}
recalling that $q_{f}$ stands for the FN charge of the fermion $f$ (cf. \cref{tab:FNcharges}).
Note that we can write the matrices in \cref{eq:mass-matrices-in-fn} with the help of diagonal matrices containing the corresponding FN suppression factors, i.e.
\begin{align}
	Q_{\bar L} = \mathrm{diag}(\epsilon^{q_{\bar L_e}},\epsilon^{q_{\bar L_\mu}},\epsilon^{q_{\bar L_\tau}}), \qquad Q_N=\mathrm{diag}(\epsilon^{q_{N_1}},\epsilon^{q_{N_2}},\epsilon^{q_{N_3}}),
\end{align}
and analogously for $Q_E$. Also, we can extract the Majorana scale $\Lambda_M$ from the bare Majorana matrix such that we have $M_N = \Lambda_M \,Y^N,$ with a coefficient matrix $Y^N$ whose entries are all close to $\mathcal{O}(1)$.
Then the mass matrices can be written as
\begin{align}\label{eq:mass-matrices-in-fn-D}
m_E = v\,{Q_L}\,Y^E\,{Q_E},	\quad m_D = v\,{Q_L}\,Y^\nu\,{Q_N},	\quad M = \Lambda_M\,{Q_N}\,Y^N\,{Q_N}.
\end{align}
Contrary to the situation in the $\nu$MSM \cite{Asaka:2005an,Shaposhnikov2005}, where one can usually choose a flavour basis in which $m_E$ and $M$ are both diagonal, we cannot do that here because each entry in the matrices above gets an individual FN suppression factor.

We begin by considering the Majorana matrix.
We are aiming for DM sterile neutrinos with late-time mass in the keV range.
From \cref{eq:initial_final_M} and \cref{fig:full-abundance-lines} we see that the early mass eigenvalues should be in the TeV range, i.e.
\begin{align}
	\tilde M^d = \tilde U^T\, M_N \, \tilde U = \tilde U^T\, \Lambda_M\,Y^N \, \tilde U \sim 10^3\,\mathrm{GeV}
\end{align} 
with an orthogonal matrix $\tilde U \sim \mathcal{O}(0.1)$ and $Y^N \sim \mathcal{O}(1)$ it follows that we should set 
\begin{align}
	\Lambda_M = 10^4\,\mathrm{GeV}.
\end{align}
We can now estimate the FN charges of the sterile neutrinos necessary to get suppressed masses in the keV range. 
The late mass eigenvalues are obtained with another orthogonal matrix $U$ as 
\begin{align}
M^d &= U^T \,Q_N (\Lambda_M\,Y^N)\,Q_N\, U	
\end{align}
Demanding that $M^d \sim 10^{-6}\,\mathrm{GeV}$ and with $U\sim\mathcal{O}(0.1)$, $Y^N\sim\mathcal{O}(1)$ and $\Lambda_M = 10^4\,\mathrm{GeV}$ we arrive at
\begin{align}
	M^d \sim 10^{-6}\,\mathrm{GeV} \quad
		\text{implying that} \quad \epsilon^{q_{N_i}+q_{N_j}} \sim 10^{-9},
\end{align}
which means that a good guess would be $q_{N_i} \sim 4$.
Here we will discuss two particular choices:
\begin{align}\label{eq:Moptions}
	Q_N =
	\begin{pmatrix}
		\epsilon^4 & & \\ & \epsilon^4 & \\ & & \epsilon^4	
	\end{pmatrix}
	\quad \text{as option 1, and}\quad
	Q_N=
	\begin{pmatrix}
		\epsilon^5 & & \\ & \epsilon^4 & \\ & & \epsilon^4	
	\end{pmatrix}
	\quad \text{as option 2}.
\end{align} 
In the case of option 1, because all sterile neutrinos have the same FN charge, meaning that $Q_N$ is proportional to the identity matrix, both the early and late mass matrices $M$ and $\tilde M$ are diagonalised by the same orthogonal matrices, or in other words, $U=\tilde U$.
This is not the case for option 2.
Up to this point we have been fixed the Majorana scale $\Lambda_M$ and have two possible choices for the FN charges of the sterile neutrinos such that we the early and late masses of the sterile neutrinos lie in the required range for successful DM production (see \cref{eq:relic-abundance,eq:y_eq} and \cref{fig:full-abundance-lines}).

Next we continue and consider the Dirac matrix $m_D$ from \cref{eq:mass-matrices-in-fn-D}.
Our goal is to choose $q_{L_i}$ and the entries of $Y^\nu$ such that the requirements from the beginning of this section are satisfied.
We start by considering the Casas-Ibarra parametrization \cite{Casas:2001sr}, which assumes the type I seesaw mechanism and parametrizes the Dirac mass matrix with the mass eigenvalues of the active and sterile neutrinos as input.
However, the standard form of the Casas-Ibarra parametrization assumes that the charged lepton mass matrix $m_E$ and the heavy neutrino mass matrix $M$ can be chosen to be diagonal - Casas and Ibarra did not have a FN embedding in mind \cite{Casas:2001sr}.
In contrary, since we are in a FN model, this is not the case.
Therefore we rederive the Casas-Ibarra parametrization using our non-diagonal matrices \cref{eq:mass-matrices-in-fn-D} and obtain
\begin{align}\label{eq:md}
	m_D = i\,V^\star \,\sqrt{m_\nu^d}\,R\, \tilde U\sqrt{\tilde{M}^d}\, \tilde U^T\,Q_N,
\end{align}
or equivalently
\begin{align}\label{eq:Ynu}
	Q_{\bar L}\,Y^\nu = \frac{i}{v}\,V^\star \,\sqrt{m_\nu^d}\,R\, \tilde U\sqrt{\tilde{M}^d}\, \tilde U^T.
\end{align}
where $R$ is any arbitrary orthogonal matrix.
Here, the matrix $V$ is the matrix that diagonalises the light neutrino mass matrix $m_\nu$, i.e.\,$m_\nu^d=V^\dagger m_\nu V$.
The freedom one has to choose $R$ reflects the fact that there are multiple possible choices for the entries of $Y^\nu$ that lead to the eigenvalues in $m_\nu^d$.
Also, recall from \cref{eq:theta12} that to compute $\theta_1^2$ we need $m_D$,
\begin{align}\label{eq:md211}
	\theta_1^2 \propto \sum_{\alpha}|(m_D)_{\alpha 1}|^2 = |m_D^\dagger\,m_D|_{11} = (Q_N\,\tilde U\,\sqrt{\tilde M^d}\, \tilde U^T\,R^\dagger \, m_\nu^d\,R\,\tilde U\,\sqrt{\tilde M^d}\,\tilde U^T\,Q_N)_{11}.
\end{align}
Concerning these last three equations the following remarks are in place:
\begin{itemize}
	\item The matrix $m_\nu^d$ contains the masses of the light neutrinos. Although the absolute scale remains to be determined, we know from oscillation data that \cite{Giganti:2017fhf}
		\begin{align}
		\Delta m_{2 1}^2 = m_2^2 - m_1^2 = 7.5\times10^{-5}\,\mathrm{eV}^2, \qquad 	|\Delta m_{3 1}^2| = |m_3^2 - m_1^2| = 2.5\times10^{-3}\,\mathrm{eV}^2.
		\end{align}
		Thus, assuming normal ordering and taking the lightest neutrino to be massless as a benchmark point, we can set
		\begin{align}\label{eq:mnu-eigenstates}
		m_1 = 0\,\mathrm{eV}, \qquad m_2 = 8.7\times10^{-3}\,\mathrm{eV}, \qquad m_3 = 5\times10^{-2}\,\mathrm{eV}.
		\end{align}
		With this, we can now consider $m_\nu^d$ as known. At the end of this section we will also investigate the case of non-vanishing $m_1$.
	\item Another ingredient is $V$, which is the matrix that diagonalises $m_\nu$.
		In the SM, where one can assume the charged lepton mass matrix to be diagonal, $V$ is  just the PMNS matrix.
		However, within a FN model $m_E$ is in general non-diagonal, but can be diagonalised by two unitary matrices, $W_L$ and $W_R$,
		\begin{align}\label{eq:mE}
			m_E^d = W_L^\dagger \,m_E\,W_R.
		\end{align}
		The PMNS matrix, whose parameters have been measured in oscillation experiments with a few percent precision \cite{Giganti:2017fhf}, is given by $ V_\mathrm{PMNS} = W_L^\dagger\,V $. Then, for a given $W_L$, we can compute $V$ by
		\begin{align}\label{eq:v_as_wl*pmns}
			V = W_L\,V_\mathrm{PMNS}.
		\end{align}
		Thus, by analysing the diagonalization pattern in the charged lepton sector we can easily obtain the matrix $V$.
	\item Notice that \cref{eq:Ynu} is independent of $Q_N$, i.e.\,it is independent of the FN charges of the sterile neutrinos.
		We can use \cref{eq:Ynu} to estimate the FN charges of the $SU(2)$ lepton doublets. By simply taking $\tilde M^d \sim 10^4\,$GeV, $Y^\nu\sim\mathcal{O}(1)$ and $R,V,\tilde U \sim \mathcal{O}(0.1)$ we conclude that $q_{\bar L_i} = 7$, or in other words
		\begin{align}
			Q_{\bar L} = \mathrm{diag}(\epsilon^7, \epsilon^7,\epsilon^7).
		\end{align}
 	\item From \cref{eq:md211} we see that $\theta_1^2$ turns out to be independent of $V$. Also, with regards to $Q_N$ it actually only depends on $(Q_N)_{11} = \epsilon^{q_{N_1}}$, i.e.\, it is independent of $q_{N_{2,3}}$.
		We can again use \cref{eq:theta12,eq:md211} to get a naive estimate of $\theta_1^2$ by taking $\tilde M^d \sim 10^4\,$GeV and $R,\tilde U \sim \mathcal{O}(0.1)$. For a DM mass of $M_1 = 7.1\,$keV we find $\theta_1^2\sim 10^{-7}$ which is not small enough to evade the $X$-Ray bound \cref{eq:xraybound} which demands $\theta_1^2\lesssim 4\cdot10^{-10}$.		
		Therefore, we have to investigate \cref{eq:md211} further to see if the freedom to choose any orthogonal matrix $R$ allows us to make $\theta_1^2$ small enough.
\end{itemize}
By now we have already specified
\begin{align}
	\Lambda_M = 10^4\,\mathrm{GeV}, \qquad Q_L=\mathrm{diag}(\epsilon^7, \epsilon^7,\epsilon^7), \qquad 
	Q_N= \begin{cases}
		\mathrm{diag}(\epsilon^4, \epsilon^4,\epsilon^4):\quad\text{option 1},\\
		\mathrm{diag}(\epsilon^5, \epsilon^4,\epsilon^4):\quad\text{option 2}.
	\end{cases} 
\end{align}
To continue and be able to properly compute $m_D$ and $\theta_1^2$ from \cref{eq:md211,eq:md}, we must make a choice to specify $V$, $\tilde U$ and $\tilde M^d$.
To determine $V$ we begin by considering the charged leptons, for which the mass eigenstates are known to be $m_E^d = \mathrm{diag}(m_e,m_\mu,m_\tau)= (0.511 \times 10^{-3},0.105, 1.776)\,\mathrm{GeV}.$
There are infinitely many non-diagonal matrices $m_E$ that have these eigenvalues.
We pick a particular choice by randomly generating two unitary matrices $W_L$ and $W_R$ and inverting \cref{eq:mE}.
From the resulting $m_E$ we see that a good choice is 
\begin{align}
	Q_E = \mathrm{diag}(\epsilon^{-3},\epsilon^{-4},\epsilon^{-4})
\end{align} in order for Yukawa matrix $Y^E$ to have $\mathcal{O}(1)$ entries, or in other words, we find that $q_E=\{-3,-4,-4\}$ is the right choice of FN charges for the charged  $SU(2)$ singlets.
In this framework, this is the origin of the hierarchy in the charged lepton masses (see the Appendix for an example).
Furthermore, having selected a specific $W_L$ we can then compute the matrix $V$ by plugging $W_L$ into \cref{eq:v_as_wl*pmns} and using the values from \cite{Giganti:2017fhf} for the PMNS matrix.
After doing so $Y^E$, $q_E$ and the matrix $V$ can be considered known.

To determine $\tilde U$ and $\tilde M^d$ we now turn to the Majorana sector, whereby we first consider option 1 from \cref{eq:Moptions}.
Notice first that because in option 1 all sterile neutrinos have the same FN charge, $Q_N$ will be proportional to the unit matrix, i.e.\,$Q_N=\epsilon^{4}\,\mathbb{1}$, and as a consequence, the late and early Majorana matrices will be related to each other simply by $M=\epsilon^{8}\,\tilde M$. 
They will also be diagonalized by the same orthogonal matrix, i.e.\,$U=\tilde U$ and their diagonal versions will also be proportional to each other, $M^d = \epsilon^{8}\,\tilde M^d$.
This means that, if we make a choice for the late masses of the sterile neutrinos $M_i$, the early masses $\tilde M_i$ are also immediately fixed.
For the DM neutrino we set $M_1=7.1\,$keV and for the other two sterile neutrinos we arbitrarily choose $M_{2}=20\,$keV and $M_{3}=30\,$keV; other choices are also possible and lead to qualitatively similar results as long as $N_2$ and $N_3$ do not contribute to the DM density (see comments below and also the right panel of \cref{fig:full-abundance-lines}).
Thus, from our choice of
\begin{align}\label{eq:eigvals-op1}
	M^d =\mathrm{diag}(7.1,\,20,\,30)\,\mathrm{keV} \qquad \text{it follows that} \qquad \tilde M^d =\mathrm{diag}(710,\,2000,\,3000)\,\mathrm{GeV}.
\end{align}
We can see from the left panel of \cref{fig:full-abundance-lines} that the lightest sterile neutrino can indeed account for 100\% of the DM if the EWPT occurs at $T_{EW} \sim 100 \,$GeV.
Furthermore, when the DM neutrino freezes-out at $T_{EW}$ and sets the DM relic abundance, the other two heavier neutrinos, which were also in thermal equilibrium with $\tilde M_{2,3}\gtrsim 2000\,$GeV, have already been depleted by their Boltzmann distributions and do not contribute to $\Omega_\mathrm{DM}h^2$, as can also be seen from the right panel of  \cref{fig:full-abundance-lines}.
Again, there are many non-diagonal Majorana matrices that have the eigenvalues specified in \cref{eq:eigvals-op1}.
We can pick a specific one by randomly generating an orthogonal matrix $\tilde U$ and computing $\tilde M = \tilde U\,\tilde M^d\,\tilde U^T$, and because $\tilde M = \Lambda_M\,Y^N$ and since we have already fixed $\Lambda_M = 10^4\,$GeV, we expect to obtain the coefficient matrix $ Y^N\sim\mathcal{O}(1)$ close to order unity.

\begin{figure}
	\begin{center}
		\includegraphics[height=7cm]{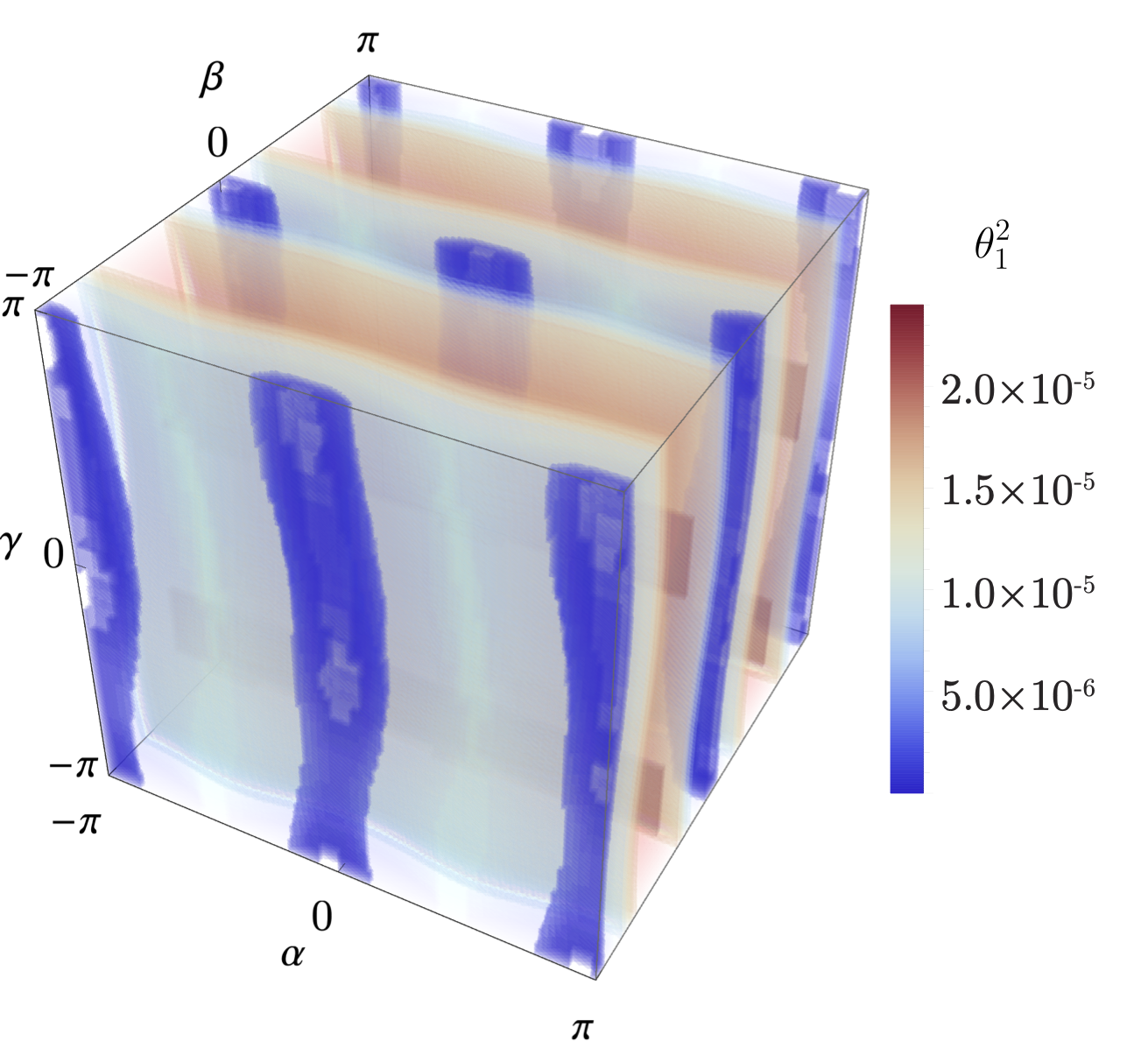}
		\caption{The mixing angle squared for the lightest sterile neutrino with all active neutrinos, $\theta_1^2$, as a function of the free parameters of the arbitrary orthogonal matrix $R$ from \cref{eq:theta-r}, i.e.\,the angles $(\alpha,\beta,\gamma)$ that parametrize $R$ as three consecutive orthogonal rotations. The smallest values are obtained in the dark blue regions of the parameter space, which seem to form three bands around $\alpha=-\pi,0,\pi$ and $\beta=-\pi,0,\pi$.}
		\label{fig:3dcube}
	\end{center}
\end{figure}

Now we finally have all ingredients to compute $\theta_1^2$, which, from \cref{eq:theta12,eq:md211}, is given by
\begin{align}\label{eq:theta-r}
\theta_1^2 = \frac{|m_D^\dagger\,m_D|_{11}}{(M_1)^2} = \frac{1}{(M_1)^2}\,(Q_N\,\tilde U\,\sqrt{\tilde M^d}\, \tilde U^T\,R^\dagger \, m_\nu^d\,R\,\tilde U\,\sqrt{\tilde M^d}\,\tilde U^T\,Q_N)_{11},
\end{align}
where the orthogonal matrix $R$ is arbitrary.
We are interested in the choices of $R$ that lead to a very small mixing angle $\theta_1^2$.
As any orthogonal matrix, $R$ has three independent parameters and one possible parametrization of it consists of the product of three $3\times3$ rotation matrices with angles $\alpha$, $\beta$ and $\gamma$ around three orthogonal directions respectively. 
Ideally, we would insert this general parametrization of $R$ in \cref{eq:theta-r} and simply compute its global minimum as a function of the three rotation angles $\alpha$, $\beta$ and $\gamma$.
However, the resulting system of equations is highly nonlinear, which makes the optimazation task difficult and inefficient. 
It is more feasible to numerically search for a local minimum, but for that we need to identify a region in the parameter space $(\alpha,\beta,\gamma)$ in the vicinity of which the minimum is expected to be found.
To graphically find such regions, we sample values of $\theta_1^2$ in a $(\alpha,\beta,\gamma)$ cube.
The result, which can be seen in \cref{fig:3dcube}, indicates the regions of the parameter space $(\alpha,\beta,\gamma)$ for which $\theta_1^2$ becomes interestingly small in dark blue color.
These regions appear to form three bands around $\alpha=-\pi,0,\pi$ and $\beta=-\pi,0,\pi$. 
The parameter $\gamma$ seems to have less of an effect.

To have a closer look we plot the plane at $\gamma=0$, which is representative for the whole cube.
\begin{figure}[bt]
\begin{center}
	\includegraphics[width=1\textwidth]{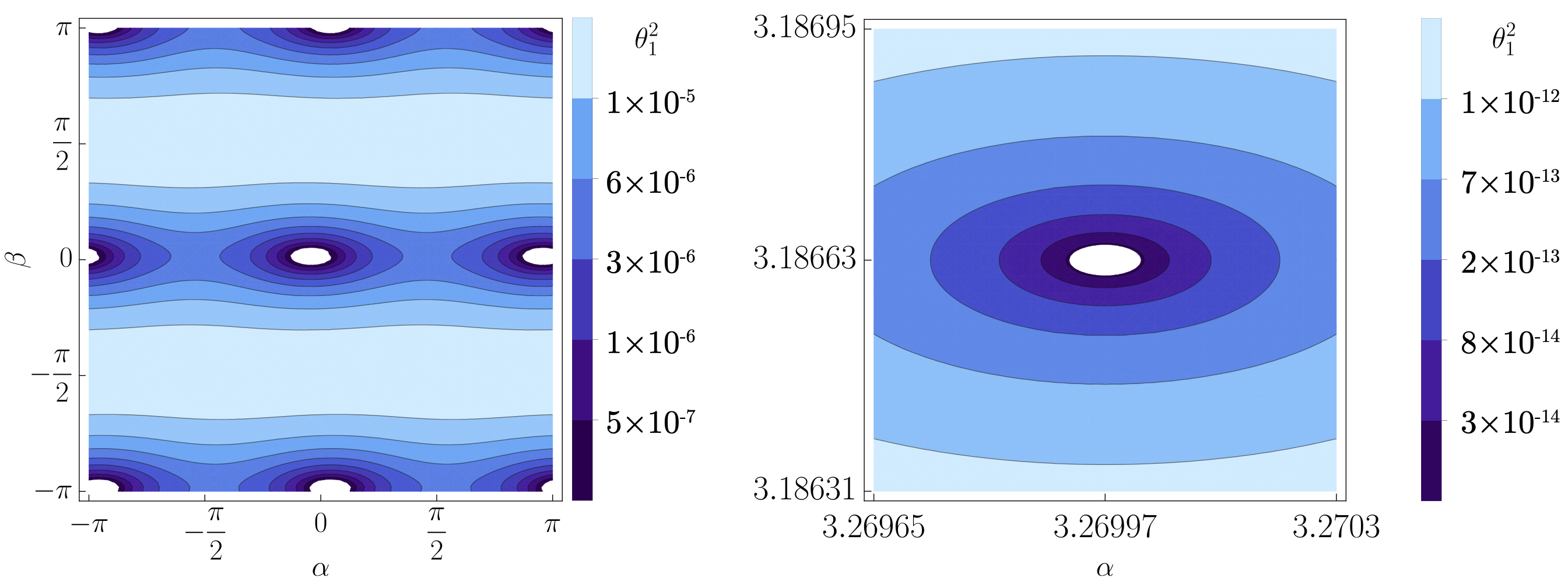}
	\caption{Left panel: contour lines for the mixing angle of the lightest sterile neutrino $N_1$ with all active neutrinos, as defined in \cref{eq:theta-r}. There are infinitely many neutrino Yukawa matrices $Y^\nu$ that could lead to the eigenvalues defined in \cref{eq:mnu-eigenstates}. The Casas-Ibarra parametrization captures this freedom in the three free parameters contained in the arbitrary orthogonal matrix $R$, which can be parametrised as a general rotation in three dimensions. Thus, $\theta_1^2$ in \cref{eq:theta-r} can be understood as a function of the three rotation angles $(\alpha,\beta,\gamma)$ contained in the matrix $R$. In order to be compatible with the $X$-Ray bounds, we want to find the matrices $R$ that make $\theta_1^2$ the smallest. In \cref{fig:3dcube} we sampled values of $\theta_1^2$ in the parameter space $(\alpha,\beta,\gamma)$. This plot is a cut of the cube in \cref{fig:3dcube} at $\gamma=0$.
	  We clearly find four different regions in the $(\alpha,\beta)$ plane where $\theta_1^2$ gets very small, potentially realising the longevity of the sterile neutrino DM (the regions of interest in the edges should not be counted twice because of the periodicity of this parametrization of $R$). As can be seen in \cref{fig:3dcube}, this cut of the cube at $\gamma=0$ is representative of the whole parameter space, and cuts at other values of $\gamma$ are qualitatively equivalent to the one presented here and have always four regions of local minima. \\ Right panel: We zoom into the local minimum of the upper right corner in the left panel of the figure and confirm that $\theta_1^2$ indeed turns small enough for the DM sterile neutrinos to be safe from the $X$-Ray constrains \cref{eq:xraybound}, which, for $M_1=7.1\,$keV demand that $\theta_1^2\lesssim 4.2\times10^{-10}$.}
	\label{fig:alpha-beta-1}
\end{center}
\end{figure}
The result is $\theta_1^2$ as a scalar field over the $(\alpha,\beta)$ plane and can be found in \cref{fig:alpha-beta-1},
where in the left panel we can see that there are four regions of interest where $\theta_1^2$ gets very small (the regions on the edges should not be counted twice because of periodicity).
As an example, we zoom into the region of interest at $(\alpha,\beta)=(\pi,\pi)$ in the right panel of \cref{fig:alpha-beta-1}.
For our DM sterile neutrino with mass $M_1=7.1\,$keV, the $X$-Ray bound of \cref{eq:xraybound} demands $\theta_1^2\lesssim 4.2\times10^{-10}$.
This bound is respected with $R(\alpha,\beta)$ for any $(\alpha,\beta)$ point in the right panel of \cref{fig:alpha-beta-1}.
In fact, the numerically determined minimum is found at $(\alpha,\beta)=(3.26997,\,3.18663)$ and has a value of $(\theta_1^2)_\mathrm{min}=1.59\times 10^{-20}$.
We emphasise that this is not the only region where we find orthogonal matrices $R$ that make our sterile neutrino DM compatible with the $X$-Ray bound from \cref{eq:xraybound}: alone in the left panel of \cref{fig:alpha-beta-1} we have three other such regions, and keep in mind that the angle $\gamma$, which is also a free parameter of $R$, was set to $\gamma = 0$ in the right panel of \cref{fig:alpha-beta-1}, but any other value of $\gamma$ would lead to qualitatively equivalent results, as we checked.
Thus, there are infinitely many regions in the parameter space of $R$ which allow our DM neutrino to evade the $X$-Ray bounds.
Furthermore, although we arrived at the result of \cref{fig:alpha-beta-1} after using one randomly generated orthogonal matrix\footnote{We also randomly generated $W_L$ to calculate $V$, but that plays no role in the computation of $\theta_1^2$; it only comes into the computation of $m_D$, see \cref{eq:md,eq:md211}.}, namely $\tilde U$, we checked that the results of this analysis are qualitatively unaltered if we use a different $\tilde U$ in \cref{eq:theta-r}.
In that case \cref{fig:3dcube} might then look slightly different, but that is irrelevant; what matters is only whether or not it is possible to find matrices $R$ such that the $X$-Ray bound is respected, and as it turns out, the answer yes in all cases.
Thus, we conclude that it is always possible to find many matrices $R$ such that our lightest sterile neutrino $N_1$ is a good DM candidate that satisfies the $X$-Ray constrain \cref{eq:xraybound}.
Also, for many of these viable choices of $R$, the neutrino Yukawa couplings turn out to be all close to order 1, as are the Yukawa matrix for the charged leptons $Y^E$ and the coefficient matrix for the Majorana mass matrix $Y^N$.
The hierarchy in the lepton masses is thus explained by interplay between the seesaw mechanism and the FN charges of the leptons.
And, since we started our computations with the Casas-Ibarra parametrization, the masses of the active neutrinos also come out correctly.

\begin{figure}[tb]
\begin{center}
	\includegraphics[width=1.1\textwidth]{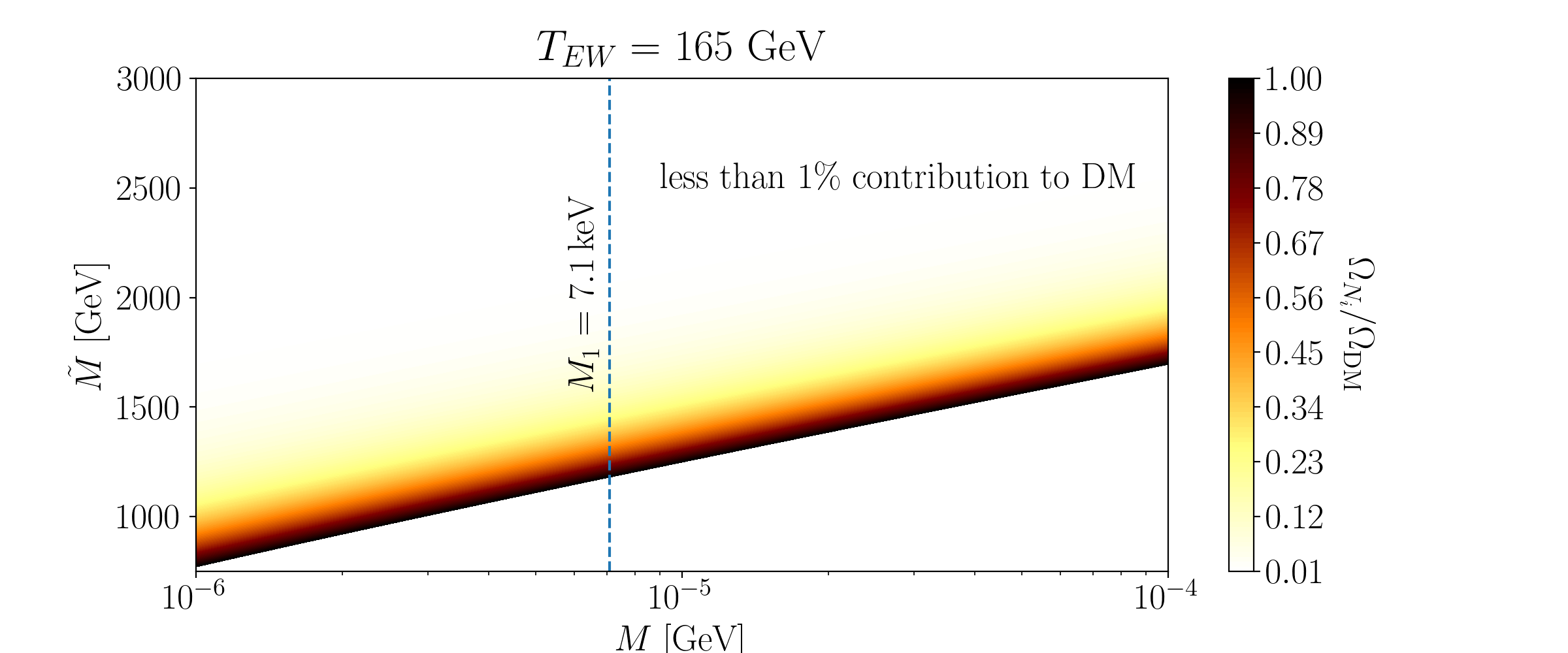}
	\caption{Contribution to the DM abundance as a function of the early and late sterile neutrino mass for freeze-out at $T_{EW}=165\,$GeV and FN charges of the sterile neutrinos given by $q_N=\{5,\,4,\,4\}$. The dashed line marks the position of our DM neutrino at $M_1=7.1\,$keV. One clearly sees that with a corresponding early mass eigenvalue of $\tilde M_1\approx 1.2\,$TeV the lightest sterile neutrino $N_1$ makes up 100\% of the DM relic abundance. The other two sterile neutrinos with late masses $M_2=20\,$keV and $M_3=30\,$keV and early masses $\tilde M_{2,3}\gtrsim 2.5\,$TeV contribute practically nothing to the DM density as their equilibrium density has already been completely depleted by their Boltzmann suppression at the moment of freeze-out.}
	\label{fig:dm-abundance-option2}
\end{center}
\end{figure}

We now make a few remarks about option 2 for the choice of FN charges of the sterile neutrinos (see \cref{eq:Moptions}), where, instead of all sterile neutrinos having the same charge of 4, we choose $q_N=\{5,\,4,\,4\}$ which is implies $Q_N=\mathrm{diag}(\epsilon^5,\, \epsilon^4, \,\epsilon^4)$.
In this case, there is an additional complication, namely that the eigenbasis of the early and late Majorana mass matrices are not the same. 
Recall that their definition is
\begin{align}
	\tilde M = \Lambda_M\,Y^N, \qquad M = \Lambda_M\,{Q_N}\,Y^N\,{Q_N}.
\end{align}
If the sterile neutrinos have different FN charges, then these two matrices are not proportional to each other, which means that they will be diagonalized by two different matrices $\tilde U \neq U$ and their diagonal versions also will not be proportional to each other.
We can still choose the late Majorana mass eigenvalues as $M^d =\mathrm{diag}(7.1,\,20,\,30 )\,\mathrm{keV}$ and use a randomly generated orthogonal matrix $U$ to find $M = U\,M^d\,U^T$.
Then to calculate $\theta_1^2$ and check for the longevity of the DM neutrino, we need to first compute the early mass matrix 
\begin{align}
	\tilde M = Q_N^{-1}\,M\,Q_N^{-1}
\end{align} 
and then compute its eigenvalues, which are the elements of $\tilde M^d$, and its normalized eigenvectors, which are the columns of $\tilde U$.
However, recall from \cref{eq:relic-abundance} and \cref{fig:full-abundance-lines} that successful DM production with $M_1=7.1\,$keV and at reasonable freeze-out temperatures for the EWPT, say $T_{EW}\in [100,\,300]\,$GeV, requires that $\tilde M_1 \in [700,\,2000]\,$GeV and $M_{2,3}\gtrsim 3000\,$GeV.
Thus, we must make sure that the early mass eigenvalues of the sterile neutrinos satisfy these conditions\footnote{When the FN charges of all sterile neutrinos are the same $q_{N_i}=q_N$, this is not necessary, because the late and early Majorana mass eigenvalues are proportional to each other as $M^d = \epsilon^{2\,q_N}\,\tilde M^d$ and we can therefore choose the late eigenvalues such that the early eigenvalues lie in the right range for successful DM genesis}, which depend on the randomly generated matrix $U$.
We checked that DM genesis is successful for $\sim 70\%$ of the randomly generated $U$ matrices, which shows that, even with this choice of FN charges for the sterile neutrinos, it is not rare to have early and late mass eigenvalues in the right range for FM production.
As a representative example, consider a case in which the early Majorana mass eigenvalues were $\tilde M^d =\mathrm{diag}(1.19,\,2.86,\,124.94)\,\mathrm{TeV}$. 
The lightest sterile neutrino contributes 100\% of the DM density for $T_{EW}\approx165\,$GeV and the other two sterile neutrinos do not contribute to it at all, as can be seen in \cref{fig:dm-abundance-option2}.
Now that we know that DM production is successful with these mass eigenvalues and freeze-out temperature $T_{EW}\approx165\,$GeV, we can then plug $\tilde M^d$ and $\tilde U$ into \cref{eq:theta-r} and repeat the previous analysis using the same parametrization of $R$ as a general rotation matrix in three dimensions to see if the DM neutrinos are sufficiently long lived.
The result looks qualitatively very similar to \cref{fig:alpha-beta-1}; there are four local minima in the vicinity of $(\alpha,\,\beta)=(0,0),\,(0,\pi),\,(\pi,0),\,(\pi,\pi)$, the value of the minima is $\theta_1^2\sim 10^{-21}$, which means that the $X$-Ray bound \cref{eq:xraybound} are satisfied within infinitely many regions of the parameter space of $R$ around the local minima.
This shows that for this new DM production mechanism to work it is not necessary that all sterile neutrinos have the same FN charges.

\begin{figure}[tb]
\begin{center}
	\includegraphics[width=0.8\textwidth]{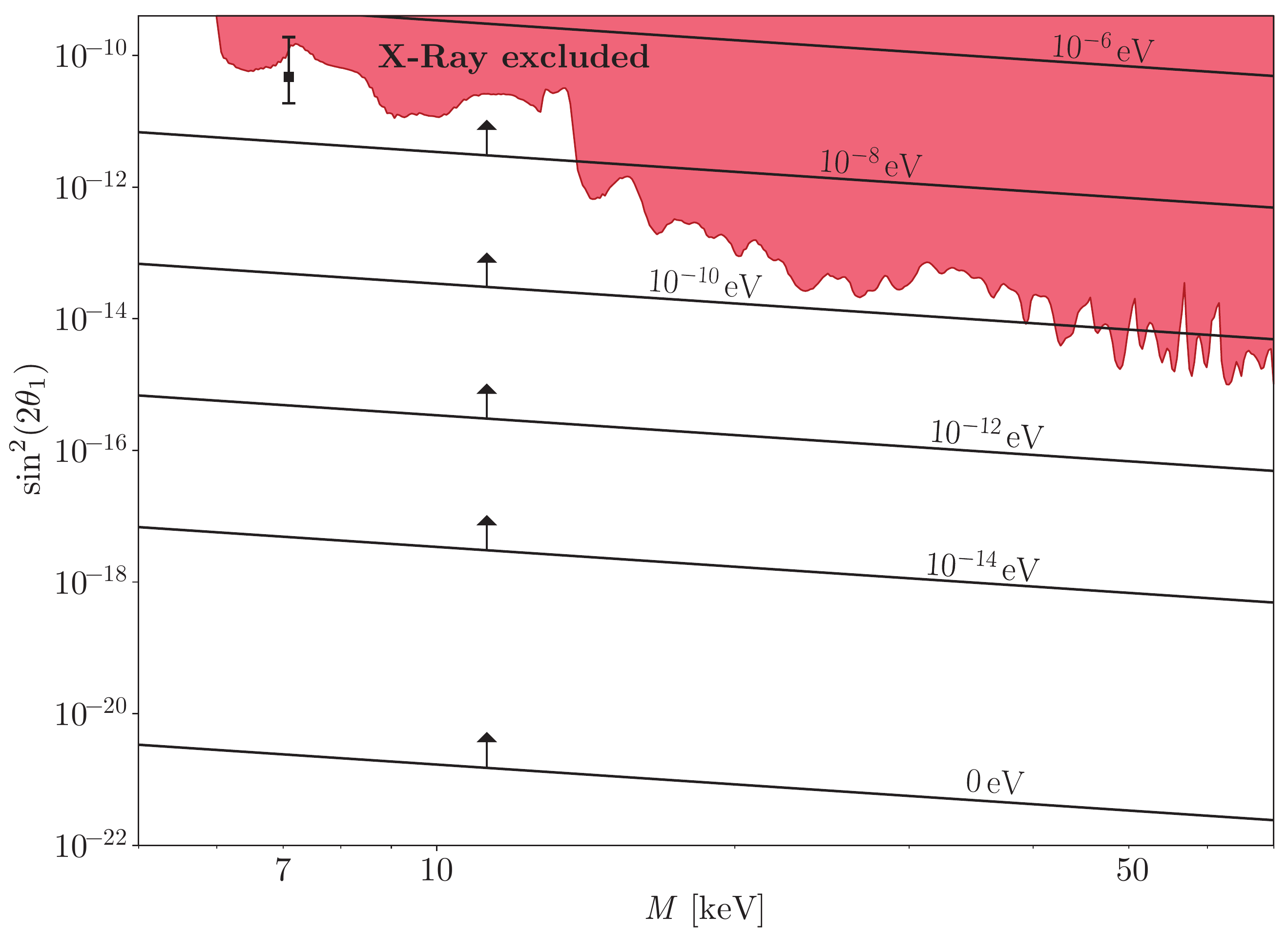}
	\caption{The parameter space for sterile neutrino DM. The red region in the upper part of the plot is excluded by the $X$-Ray survey NuSTAR GC (2016) \cite{Perez:2016tcq}. In the framework presented in this work, how far down the mixing angle can be pushed depends on the absolute mass scale of the active neutrinos. Each solid line gives, for the denoted mass of the lightest active neutrino, the smallest mixing angle that could be reached with the given DM mass, while still recovering the known squared mass differences and oscillation data of the active neutrinos. Thus, the area above each line is the region of the parameter space where the framework proposed here gives us a viable DM candidate, the lightest sterile neutrino produced by the thermal freeze-out induced during the EWPT. If the lightest active neutrino is massless, then $\sin^2(2\theta_1)$ can be as small as $10^{-21}$ and safely evade all $X$-Ray bounds.}
	\label{fig:m-t-params-space}
\end{center}
\end{figure}

Up to this point we have focused on a DM neutrino with mass $M_1=7.1\,$keV, which is particularly interesting because a suspicious $X$-Ray signal has been detected \cite{Drewes2012}.
Now we turn to the possibility of DM neutrinos with masses different than $M_1=7.1\,$keV.
For this it is best to return to the previous choice of FN charges for the sterile neutrinos, i.e.\,option 1 from \cref{eq:Moptions}, where all three sterile neutrinos have the same FN charge $q_N=\{4,\,4,\,4\}$. We will soon see why this is convenient.
Our starting point is the same choice of late sterile neutrino mass eigenvalues with which we have been working all along, namely $M^d=\mathrm{diag}(7.1,\,20,\,30)\,$keV.
The corresponding early mass eigenvalues are $\tilde M^d=\mathrm{diag}(0.71,\,2,\,3)\,$TeV.
By multiplying the late mass eigenvalues with a dimensionless parameter $s$ and letting $s$ vary, we can homogeneously vary all late and early mass eigenvalues, because with the current choice of FN charges $M^d$ and $\tilde M^d$ are proportional to each other.
Of course, for $s=1$ we recover the case that we have been studying so far.
Next we determine how $\theta_1^2$ changes when the sterile neutrino mass eigenvalues vary as determined by the dimensionless parameter $s$.
From \cref{eq:theta-r} we see that $\theta_1^2$ is proportional to two powers of $\sqrt{\tilde M^d}$ while being inversely proportional to $(M_1)^2$.
Consequently, the dependence is $\theta_1^2 \propto s^{-1}$.
This allows us to extrapolate from the point $(M_1 = 7.1\,\mathrm{keV},\,\theta_1^2\approx10^{-21})$ to span a wide area of the $(M_1,\,\theta_1^2)$ plane,
the results of which can be seen in \cref{fig:m-t-params-space}.
We checked that the extrapolation delivers correct results by explicitly computing $\theta_1^2$ for different combinations of late and early mass eigenvalues.
At this point we also relax the fix-point assumption that one of the active neutrinos is massless and compute one point in the parameter space to extrapolate from for non-vanishing values of $m_1$.
We find that the ability of this DM candidate to evade the $X$-Ray bounds relies on the lightest active neutrino being massless or at least much lighter than the other two active neutrinos.
In \cref{fig:m-t-params-space} each solid black stands for a different mass for the lightest active neutrino, with the line at the bottom representing the case of one massless active neutrino.
The region above each line is the portion of the parameter space where the FN model and DM production mechanism proposed here delivers a viable DM candidate, the lightest sterile neutrino, which is stable on cosmological timescales and evades the $X$-Ray constrains. 
Furthermore, for $s=7.05$ the late DM mass reaches $M_1=50\,$keV while the corresponding early mass reaches slightly over $\tilde M_1=5\,$TeV.
For this combination of masses to achieve successful DM production according to \cref{eq:relic-abundance}, the freeze-out temperature would need to be $T_{EW}\approx550\,$GeV, which is not unreasonable. 

\section{Conclusions}
\label{sec:conclusio}

In this work, we have proposed a new mechanism for keV sterile neutrino DM genesis in the early Universe that relies neither on the oscillations between active and sterile neutrinos nor on heavier parent particles which produce the DM sterile neutrinos in their decay.
Instead, the sterile neutrinos are produced by freeze-out from thermal equilibrium, much like a typical WIMP, without introducing any additional coupling or interactions for the sterile neutrinos beyond their Yukawa term, by which they generate the masses of the light neutrinos through the type I seesaw mechanism, and the FN interactions encoded by the flavon and visible only in the UV-complete theory.
This possibility has not been considered in the past because for the sterile neutrinos to be in thermal equilibrium their Yukawa couplings would need to be sizeable, which would compromise their stability.
Indeed, the decay of sterile neutrinos with a non-vanishing Yukawa coupling is unavoidable and has been searched for with $X$-Ray surveys, placing strict upper bounds on the mixing angle between sterile and active neutrinos, and equivalently, on the Yukawa coupling.
As we pointed out previously in \cite{Jaramillo_2021}, this can be solved by varying Yukawa couplings: if, by some mechanism, the Yukawa couplings were large at early times but became very small at later times, then the sterile neutrinos could have been in thermal equilibrium in the early Universe, then decouple and stay stable on cosmological timescales thereafter.

Here we have presented an implementation of varying Yukawa couplings by formulating a FN model for the leptonic sector expanded by three sterile Majorana neutrinos.
In this framework the Yukawa couplings and the Majorana mass matrix of the sterile neutrinos are rescaled powers of the FN suppression factor, which is proportional to the vev of the flavon.
We argue that during the Electroweak phase transition, when the vev of the Higgs changes from 0 to $v$ as the scalar potential finds its true minimum, it is reasonable to expect that the vev of the flavon will change too.
We have shown that, assuming for concreteness that the FN factor was close to 1 before the Electroweak phase transition and 0.1 afterwards, it is possible for the sterile neutrinos to be produced by thermal freeze-out and later evade the current $X$-Ray constrains.
Because of their unsuppressed Yukawa couplings, the sterile neutrinos, like all other fermions, are in thermal equilibrium before the Electroweak phase transition.
When the phase transitions occurs and the vev's of the Higgs and the flavon get their late-time values, the Yukawa couplings get suppressed, and in the case of the sterile neutrinos this suppression is so drastic that they are forced to decouple from the thermal bath and are able to stay compatible with the $X$-Ray constrains.
In this sense it is the phase transition itself which induces the decoupling of the sterile neutrinos.
Furthermore, the change in the FN factor during the phase transition also provokes a change in the masses of the sterile neutrinos.
A consequence of this is the fact that reproducing the observed relic abundance by this induced freeze-out mechanism relies on the interplay between the early- and late masses of the sterile neutrinos as well as on the temperature of the phase transition, which is also the temperature of freeze-out.
For DM masses in the keV range, the Majorana scale and the masses of the sterile neutrinos in the early Universe should be on the TeV scale and the temperature of the phase transition should lie between $100-600\,$GeV.

In the specific FN realization proposed here, we show that successful DM genesis is possible while also obtaining Yukawa couplings and entries in the Majorana coefficient matrix that are all of order unity or close to it, thus somewhat alleviating the flavour puzzle in the lepton sector by the different FN charges of the leptons, and avoiding a severe hierarchy in the Yukawa couplings. 
The smallness of the active neutrino masses and oscillation phenomena are explained by the synergy of the FN model and the type I seesaw mechanism.
We empirically find that the smallest value of the mixing angle of the DM neutrino with all active neutrinos that can be achieved in this framework depends of the absolute scale of the active neutrino masses, and is minimal in the case of one massless active neutrino. 

At this point, one remark concerning the spectrum of the sterile neutrino dark matter produced by this mechanism is in place.
Right until the moment of induced freeze-out by the phase transition, the sterile neutrinos have a non-relativistic thermal spectrum, i.e.\,it would be cold DM.
The question of whether this changes after the phase transition due to the effective suppression of the Majorana masses is non-trivial and will be connected to the dynamics of the phase transition.
In Ref. \cite{Baldes:2016rqn} it has been discussed that larger Yukawa couplings could lead to the EWPT becoming a first order phase transition, for which the friction between the bubble wall and the plasma plays an important role in the wall dynamics, among other things \cite{Espinosa:2010hh}.
In essence, the question at hand boils down to how the coupled particles exchange energy and momentum during the phase transition across the bubble walls.
A detailed analysis of this issue lies beyond the scope of this study and we leave it for future work.

In summary, with only three additional sterile neutrinos and one flavon (which encodes the FN messengers that are integrated out at low energies - of course) the framework proposed here allows us to alleviate the flavour problem, the origin of the masses of active neutrinos and their oscillation phenomena and produces a viable and appealing DM candidate. 
Future investigations should look into the possibility of also addressing the baryogenesis/leptogenesis problem and explore other possible implementations of varying Yukawa couplings to achieve the same goals.

\begin{center}
\textbf{Acknowledgements}\\
\end{center}
I gratefully thank Dr.\;Salvador Centelles Chulia, Dr.\;Sudip Jana and Dr. Andreas Trautner for many invaluable discussions  and feedback as well as Prof. Dr. Manfred Lindner for his support.
\appendix
\section{One specific benchmark point in the FN model}\label{sec:apdxA}
We start with the following mass eigenvalues for the SM leptons:
\begin{align}
	m_E^d &= \mathrm{diag}(0.511 \times 10^{-3},\, 0.105,\, 1.776)\,\mathrm{GeV}, \\
	m_\nu^d &= \mathrm{diag}(0,\, 8.7\times10^{-3},\, 5\times10^{-2})\,\mathrm{eV},
\end{align}
and for the mass of the sterile neutrinos before the EWPT we choose
\begin{align}
	\tilde M^d &= \mathrm{diag}(710,\, 2000,\, 3000)\,\mathrm{GeV}. 
\end{align}
The FN charges for the charged singlets, the doublets, and the sterile neutrinos are taken to be
\begin{align}
	q_E = \{-3,\,-4,\,-4\}, \qquad q_L = \{7,\,7,\,7\}, \qquad q_N = \{4,\,4,\,4\}.
\end{align}
respectively. The bare Majorana mass matrix is given by $\tilde M = \Lambda_M\,Y^N$, where we set $\Lambda_M=10^4\,$GeV, and is related to its diagonal version by the orthogonal matrix $\tilde U$, which we set to
\begin{align}
	\tilde U = \begin{pmatrix}
		-0.200 & -0.696 & -0.690\\
 0.868 &  0.209 & -0.460\\
-0.465 &  0.687 & -0.559
	\end{pmatrix}, \quad \text{leading to} \quad 
	Y^N = \begin{pmatrix}
		0.242 & 0.054 & 0.027\\
0.054 & 0.125 & 0.077\\
0.027 & 0.077 & 0.203
	\end{pmatrix}
\end{align}
After the EWPT, the suppressed mass eigenvalues of the sterile neutrinos are
\begin{align}
	M^d &= \mathrm{diag}(7.1,\, 20,\, 30)\,\mathrm{keV}.
\end{align}
Since in this case all sterile neutrinos have the same FN charge, the eigenbasis of the matrices $\tilde M$ and $M$ are the same.
The randomly generated unitary matrix $W_L$, used to diagonalize the charged lepton mass matrix   by $m_E^d = W_L^\dagger \,m_E\,W_R$ and compute the matrix $V = W_L\,V_\mathrm{PMNS}$ used in \cref{eq:md} is given by
\begin{align}
	W_L = \begin{pmatrix}
		-0.598 + i\,0.289 & -0.508 - i\,0.023 & -0.548 + i\,0.009 \\
-0.377 + i\,0.531 &  0.558 + i\,0.123 &  0.161 - i\,0.474 \\
-0.364 - i\,0.057 &  0.485 - i\,0.424 & -0.054 + i\,0.668 
	\end{pmatrix}.
\end{align}
Finally, the Yukawa matrices for the neutrinos and charged singlets are
\begin{align}
	|Y^\nu| = \begin{pmatrix}
		0.030 & 7.381 & 1.031\\
		0.029 & 4.681 & 4.068\\
		0.025 & 3.308 & 4.142
		\end{pmatrix},
		\qquad
		|Y^E| = \begin{pmatrix}
		3.14 & 2.59 & 2.98\\
		1.89 & 2.21 & 2.86\\
		5.57 & 3.28 & 3.53
		\end{pmatrix}.
\end{align}
Although there still is some hierarchy in the columns of $Y^\nu$, it is remarkably less severe. 
The mixing angle for the DM neutrino with all active neutrinos is $\theta_1^2 = 2.86\times 10^{-10}$ which is below the bound of $\theta_{1\,\mathrm{bound}}^2 = 4.16\times10^{-10}$ at a DM mass $M_1=7.1\,$keV.

\newpage
\bibliographystyle{unsrturl}
\bibliography{/Users/charlie/MPI_cloud/00_Forschung/01_references/central_library}

\end{document}